\documentstyle[preprint,revtex]{aps}
\begin{document}
\draft
\begin{title}
Relativistic Nuclear Matter with Self-Consistent Correlation Energy
\end{title}

\author{J. A. McNeil}
\begin{instit}
Physics Department, Colorado School of Mines, Golden, Colorado 80401
\end{instit}

\moreauthors{C. E. Price and J. R. Shepard}
\begin{instit}
Physics Department, University of Colorado, Boulder, Colorado 80301
\end{instit}

\begin{abstract}
We study relativistic nuclear matter in the $\sigma - \omega$ model

including the ring-sum correlation energy. The model parameters are

adjusted self-consistently to give the canonical saturation density
and

binding energy per nucleon with the ring energy included. Two models

are considered,  mean-field-theory where we neglect vacuum effects,
and

the relativistic Hartree approximation where such effects are
included

but in an approximate way. In both cases we find self-consistent

solutions and present equations of state. In the mean-field case the

ring energy completely dominates the attractive part of the energy

density and the elegant saturation mechanism of the standard approach

is lost, namely  relativistic quenching of the scalar attraction.  In

the relativistic Hartree approach the vacuum effects are included in
an

approximate manner using vertex form factors with a cutoff of 1 - 2

GeV, the range expected from QCD. Due to the cutoff, the ring energy

for this case is significantly smaller, and we obtain self-consistent

solutions which preserve the basic saturation mechanism of the
standard

relativistic approach.

\end{abstract}

\section{Introduction}

Relativistic models of nuclei based on meson-nucleon degrees of
freedom

have a long history. One particularly elegant class of relativistic

models, quantum hadrodynamics (QHD), has been developed by Walecka
and

others \cite{serot}.  In its simplest manifestation (QHD-I) isoscalar

Lorentz-scalar ($\sigma$) and isoscalar Lorentz-vector ($\omega$)

mesons are coupled to nucleon fields.  With this renormalizable (but

not asymptotically free) theory one has the means to include vacuum

effects in a well-defined (if perhaps physically dubious) way.

Typically the modelis solved in mean-field-theory (MFT) and the

parameters are fit to bulk nuclear properties. Such models have been

applied successfully to a wide variety of nuclear phenomena from
static

properties to dynamic nuclear responses
\cite{horowitz}--\cite{mcneil}.

One of the more appealing features of these models (at least in its

mean field manifestation) is its elegant saturation mechanism, namely

the relativistic quenching of the scalar attraction at high
densities.

Attempts to go beyond the Hartree (one-loop) level have proved

difficult.  The loop expansion appears not to converge as the

self-consistent 2-loop solutions are radically different from their

one-loop cousins \cite{2loop}.  In a preliminary effort to examine

alternative organization schemes we have calculated the correlation

energy in the ring sum approximation based on the mean field

(no-vacuum) basis. (This quantity was calculated previously by Ji

\cite{ji}.)  Using mean field parameters we confirm Ji's calculation
of

a large correlation energy, -33 MeV per nucleon. Thus, the
correlation

energy is large compared to the mean field energy and may not be

treated as a perturbation. In this work we include the correlation

energy self-consistently in fixing the model parameters.  We find a

stable solution set and present the resulting equation of state.  We

also calculate the compressibility, examine the saturation mechanism,

and compare to the standard mean field results.

The role of the vacuum in relativistic nuclear models based on

meson-nucleon degrees of freedom is controversial \cite{cohen}.  We

know that nucleons have substructure on the scale of $\sim 1$ GeV.
For

QHD-I at the one-loop level energy scales beyond this are
significant,

but do not dominate the physics at least as regards the saturation

mechanism and the low energy phenomenology.  The relativistic Hartree

results are qualitatively the same and quantitatively similar to the

mean field results. Beyond one-loop the high energy sector completely

dominates the physics in an uncontrolled manner.  Prakash, et al.
argue

that the subnucleon structure implies a cutoff on scales of 1 GeV.

They calculate the two-loop contributions including vertex cutoffs on

this scale and found that the two-loop terms were small
\cite{prakash}.

In addition it has been shown that QHD is incompatible with the large

$N_c$-limit of quantum chromodynamics (QCD) thought to be the basis
of

the strong interaction \cite{qcd}. To mimic the softening of the
vertices

in the ring energy due to the onset of subnucleonic degrees of
freedom

which are beyond the scope of the present model, we calculate the
vacuum

polarized correlation energy using renormalized vacuum polarizations
with

vertices regularized by vertex form-factors.  In this way we retain
the

``low-lying'' effects of the vacuum.  The price we pay is that the
results

are quantitatively sensitive to the choice of cutoff parameter in the

vertex form factor. Nevertheless, we know that the scale of such
effects

is about 1 GeV, and we find that the qualitative features of the
results

are similar for any cut-off on this scale.  Using this regularization

scheme, we self-consistently fit the model parameters including the

correlation energy.

\section{Correlation Energy}

We begin with a review of the mean field theory description of nuclei

based on the QHD-I model of Walecka \cite{serot}.  This particularly
simple

yet efficient model consists of three fields, the nucleon ,$\psi$,

the scalar-isoscalar sigma, $\phi$, and the vector-isoscalar omega,

$V^{\mu}$. The Lagrangian density is given by

\begin{eqnarray}
{\cal L}(\psi ,\phi , V^{\mu} )&= \bar{\psi}[\gamma_{\mu}
(i\partial^{\mu}&-g_vV^{\mu})-(M-g_s\phi)]\psi

+{1\over 2}(\partial_\mu\phi\partial^\mu\phi-m_s^2\phi^2) \nonumber
\\
&&-{1\over 4}F^{\mu\nu}F_{\mu\nu}+{1\over 2} m_v^2V_\mu V^\mu

\label{lag}
\end{eqnarray}
where $F_{\mu \nu} = \partial_{\mu} V_{\nu}-\partial_{\nu}V_{\mu}$.
A

standard counter term is added when the vacuum is considered. In this
work

we neglect Coulomb and isovector interactions. We will also ignore

exchange terms throughout as the proper inclusion of such terms
requires a

treatment of the isovector mesons as well.

We wish to focus our attention on the ring-sum polarization energy
density

given by

\begin{equation}
{\cal E}_{ring}(k_F,M^*)=-{i\over 2}\int{d^4q\over (2\pi)^4}\{

{\rm Tr}(D\Pi_0)+\ln [\det (1-D\Pi_0)]\}
\end{equation}
where $\Pi_0$ is the mixed scalar-vector polarization insertion which
can

be represented by a 3x3 scalar-vector  matrix and a 2x2 (diagonal)

transverse vector matrix \cite{chin}:

\begin{eqnarray}
\Pi_{sv} \: &=& \: \left(

      \begin{array}{ccc}

                    q^2 \Pi_l & wq\Pi_l & \Pi_m     \\
                    -wq\Pi_l   &-w^2\Pi_l & -w/q\Pi_m \\
                    \Pi_m     &w/q\Pi_m &  \Pi_s

       \end{array} \;

		    \right)

\end{eqnarray}

\begin{eqnarray}
\Pi_{trans} \: &=& \:\left( \;
                    \begin{array}{cc}
                         \Pi_T & 0     \\
                          0    & \Pi_T

		     \end{array}\;

	      \right)
\end{eqnarray}
where the 4-momentum transfer is $q^{\mu}= (\omega,\vec q)$. Kurasawa
and

Suzuki\cite{kurasawa} have provided analytic forms for the various

polarizations. The use of analytic forms greatly reduces the
computing

time.

The analytic structure of the integrand allows a Wick rotation.
Defining

$\omega = iq_0 $, we have
\begin{equation}
q_{\mu}^2 = \omega^2-q^2 = -(q_0^2+q^2)=-q_E^2
\end{equation}
where $q_E$ is the Euclidean 4-momentum.  The integrals then become
\begin{equation}
{\cal E}_{ring}(k_F,M^*) ={1\over (2\pi)^3}\int_0^{\infty}dq_Eq_E^3
                              \int_0^{2\pi}d\theta_E cos^2(\theta_E)
                               \{Tr(D\Pi_0)+ln[det(1-D\Pi_0)]\}
\end{equation}
where $q_0 = q_Esin(\theta_E), q = q_E cos(\theta_E)$. The remaining
two

integrals are performed numerically.  For mean field parameters we
find

convergence at an upper limit for $q_E$  $\sim 10k_F$.  Using the
mean

field parameters given in Table I, we find a large ring energy of -33
MeV

per nucleon as did Ji \cite{ji}. It is therefore clear that within
this

model the ring-sum correlation energy is not a small effect being
twice

the magnitude of the MFT binding energy.

The significance of the ring-sum energy can be placed in a more
general

context by considering the recent lattice QHD calculations of
Brockmann

and Frank \cite{brockmann}. They claim that the effect of all quantum

corrections to the MFT is to change the binding energy by about -85
MeV

per nucleon using the canonical MFT parameters.  The ring sum
provides an

approximation to such quantum corrections.  Thus, the ``good news''
is

that the ring sum contribution - which is calculationally tractable -

represents a substantial ($\sim$40\%) fraction of the total from
diagrams

beyond MFT.  The ``bad news'' is that the remaining $\sim$60\% is due
to

much more complicated processes whose treatment by conventional
many-body

means may not be feasible.

\section{Nuclear Matter with Self-consistent Ring Energy}

We sought to include the ring-sum correlation energy in the total
nuclear

matter energy density and re-fit the coupling parameters to
saturation

 density and binding energy per nucleon. We present results where the

 meson masses were held fixed at  $m_v= 783$ MeV and $m_s = 500$ MeV.
We

 discuss the effect of varying the sigma mass later. The total energy

 density is given by

\begin{equation}
 {\cal E} = {\cal E}_{mesons}+{\cal E}_F+{\cal E}_{ring}+{\cal
E}_{vac}
\end{equation}
where
\begin{equation}
{\cal E}_F = {\gamma \over 16 \pi^2}[2k_FE_F^3-{M^*}^2k_FE_F-{M^*}^4
  {\rm ln}({k_F+E_F\over M^*})],
\end{equation}
\begin{equation}
{\cal E}_{mesons}={1\over 2}{g_v^2\over m_v^2}\rho_B^2
                 +{1\over 2}{g_s^2\over m_s^2}\rho_s^2,
\end{equation}
\begin{equation}
{\cal E}_{vac}=-{\gamma\over 16 \pi^2}[{M^*}^4ln({M^*\over M})
     -{25\over 12}{M^*}^4
      +4{M^*}^3M-3{M^*}^2M^2 +{4\over 3}M^*M^3-{1\over 4}M^4].
\end{equation}

We fix the saturation density appropriate to $k_F =1.42$ fm$^{-1}$.
The

saturation condition is
\begin{equation}
{\partial {\cal E}\over \partial k_F}\Biggm\vert_{k_f=1.42 fm^{-1}} =
0.
\end{equation}

At each step of the fitting procedure the self-consistency condition,
\begin{equation}
 M^* = M - {g_s^2\over m_s^2} \rho_s,
\end{equation}
 is maintained.

The scalar density in turn can be found from the energy density
through

the minimization condition,
\begin{equation}
{\partial {\cal E}\over \partial M^*}\Biggm\vert_{k_f=1.42 fm^{-1}} =
0,
\end{equation}
which gives
\begin{equation}
\rho_s = {\partial( {\cal E}_F+{\cal E}_{vac}+{\cal E}_{ring}) \over

\partial M^*}.
\end{equation}

As usual the Fermi sea contribution to the scalar density is

\begin{equation}
{\rho_s}^F={\partial {\cal E}_F\over \partial M^*}
     ={\gamma \over 4 \pi^2}M^* [k_FE_F-{M^*}^2\ln({k_F+E_F\over
M^*})]
\end{equation}
with the ring contribution calculated numerically.

 The coupling constants were varied until a self-consistent solution
with
 a binding energy per nucleon of 15.75 MeV and a saturation density
of

 corresponding to $k_F=1.42 fm^{-1}$ was

obtained.  Once the coupling parameters were found, the equation of
state

was generated by finding the self-consistent total energy at various

values of $k_F$ near the saturation value keeping the coupling
parameters

fixed.  Three models were considered: the mean field model and the

relativistic Hartree model with vertex cutoffs of 1 GeV and 2 GeV.
The

resulting coupling constants and bulk properties are given in Table
I.

 The resulting equation of state for the mean field model is shown in
Fig.

 1.  We found a compressibility of 304 MeV which is significantly
smaller

than the standard MFT result of around 400 - 500 MeV, and is nearer
the

values derived from analyses of the breathing modes of heavy nuclei.
Also

shown in Fig. 1 are the separate contributions from the MFT-terms and
the

correlation energy.  We see that the saturation curve is a sensitive

cancellation of the large repulsive MFT terms with the large
attractive

ring-sum energy. (At the saturation point the ring energy is -58 MeV
!).

The saturation mechanism is therefore quite different than the usual
one

of relativistic quenching of the scalar attraction with higher
density.

Here the attractive ring energy dominates at moderate densities with
the

repulsive vector-MFT term dominating at high densities. It is an open

question whether this saturation mechanism is realistic in that it is

consistent with other low energy finite nuclear phenomenology.

The ratio $M^*/M$ was found to be $\sim .8$ which is significantly
larger than the standard mean field result of $\sim .6$. With the
relatively large value of

$M^*/M$, one suspects that the spin-orbit splitting in finite nuclei
will be too small.  This issue will be explored in subsequent work.

The corresponding ring-sum energy when the vacuum is included is
divergent

even when the polarizations themselves have been renormalized because
the additional loop integral diverges.  Since the

theory is renormalizable, the new divergence may be rendered finite
through standard renormalization methods;

however this approach is not realistic in that the high energy sector

of this model (QHD) does not incorporate the physically correct
internal nucleon

degrees of freedom.  There are other ways of moderating the high
energy

sector within the scope of the QHD model alone.  Allendes and Serot
have studied effective form factors

induced in the vector meson propagator due to soft brehmstrahlung

processes\cite{allendes}. They find that the large space-like
momentum region

is quenched on the scale of 4 - 5 nucleon masses strictly within

the QHD model. However, their approach also does not incorporate the
subnucleon

degrees of freedom which are expected to dominate at scales
significantly smaller,

namely 1 GeV. As an alternative approach, we have attempted to
account for the

internal nucleon degrees of freedom in an approximate way by using
vertex form-factors to moderate the high energy sector at a
physically reasonable scale. We chose a standard dipole form:

\begin{equation}
f(q_{\mu}^2,\Lambda) ={1\over (1+q_{\mu}^2/\Lambda^2)^2}
\end{equation}
where $\Lambda$ is chosen to be of the order of 1 GeV.  Over this
scale

($1\rightarrow 2$ GeV) the results are quantitatively sensitive to

$\Lambda$ and an independent means of determining the form and scale
of

the cutoff from nucleon models is desirable.

In Fig. 2 we show the equation of state for $\Lambda = 1$ GeV and in
Fig.

3 the corresponding curve for $\Lambda = 2$ GeV. As with the previous
mean

field case the $\sigma$ mass has been fixed at 500 MeV.  For $\Lambda
= 1$

GeV the standard relativistic Hartree energy dominates.  At the
saturation

density the ring energy is just -3.4 MeV per nucleon. The
self-consistent

coupling parameters are given in Table I.  The compressibility is 503
MeV

which is close to the usual RHA result of around 450

MeV.  The ratio, $M^*/M$, is .722

which is similarly close to the usual RHA result and will probably
imply

a reasonable spin-orbit splitting in finite nuclei.  While these

quantitative values are not meaningful due to their sensitivity to
the

cutoff, we find that the fundamental saturation mechanism of the

underlying RHA is preserved and the correlation energy is a
relatively

small correction comparable to that found in non-relativistic nuclear

structure models.  For $\Lambda = 2$ GeV the correlation energy is

naturally larger. At saturation density the ring energy is -12.5 MeV.
The

coupling parameters for this case are also found in Table I. The

compressibility is 521 MeV while the ratio  $M^*/M$ is about .724.
Thus we see

that these bulk quantities are relatively insensitive to the cutoff
in the

region of 1-2 GeV.  Comparing Figures 2 and 3 one can see the trend
in the

increasing importance of the ring energy as $\Lambda$ is increased.
The

attraction from the $\sigma$ meson plays a smaller role as the
correlation

energy begins to dominate the energy in the equation of state near

saturation.

We have also solved (but not shown) the self-consistent relativistic

Hartree model for a sigma mass of 600 MeV. For the 2 GeV cutoff, we
found that the qualitative

features remain similar, but that the compressibility increases

 to around 600 MeV while the $M^*/M$-ratio drops slightly to around

.71.  The higher compressibility  results from a more rapid change in
the

ring energy with respect to $k_F$ near the saturation point for this
value

of the sigma mass as compared to the previous case. The smaller
$M^*/M$

should give about the correct spin-orbit splitting in finite nuclei.
There

appears to be a correlation between the larger compressibility and
the

smaller $M^*/M$.  This correlation was shown in previous work by

Furnstahl, Price, and Walker \cite{price2} in the context of
non-linear

QHD models, and may be a universal feature of non-linear
self-consistent

relativistic models of nuclear matter.

\section{Conclusions}

The lesson learned from this study is that relativistic nuclear
models,

such as QHD, based only on hadronic degrees of freedom are capable of

realizing a credible (i.e. successful and convergent) phenomenology
if the

high energy sector is moderated at  scales beyond 1-2 GeV.  It
remains for

quark-nucleon models to provide a quantitative understanding of the
nature

and scale of such modifications.  However, given the phenomenological

nucleon form factors, one would expect any successful quark model of

nucleons to result in effective vertex form factors with cutoffs on
the

scale of 1 GeV where, as we have shown here, models based on
meson-nucleon

degrees of freedom preserve the simple relativistic saturation
mechanism

when ring energy contributions are included. It remains to
investigate

whether other low energy phenomena can be consistently described in
this

extended model.

\section{Acknowledgements}
This work was supported in part by the National Science Foundation
and the

Department of Energy.

\newpage
\figure{ Equation of state for mean field theory with ring-sum
correlations.}

\figure{  Equation of state for relativistic Hartree theory with
ring-sum

correlations with vertex cutoff of 1 GeV.}

\figure{  Equation of state for relativistic Hartree theory with
ring-sum

correlations with vertex cutoff of 2 GeV.}

\newpage
\begin{table}
\caption{Parameters for relativistic nuclear matter fit to saturation

density ($k_F$=1.42 fm$^{-1}$) and binding energy per nucleon (-15.75

MeV). The meson masses were fixed at $m_v$=783 MeV and $m_s$=500
MeV.}

\label{parameters}
\begin{tabular}{ccccccc}
Model        & $g_s^2$ & $g_v^2$ & $M^*/M$ & $\kappa$ (MeV) &
$E_{ring}$(MeV)

& $\Lambda$ (GeV)\\
\tableline
MFT(w/o fit) & 109     & 189     &  .54          &   450     &  -33
&

-- \\
MFT(w/ fit)  & 39.6    & 104.7   &  .791         &   304     &  -58.4
&

--  \\
RHA-1        & 55.1    & 88.5    &  .722         &   503     & -3.4
&

1.0 \\
RHA-2        & 54.5    & 94.9    &  .724         &   521     &  -12.5
&

2.0  \\
\end{tabular}
\end{table}


\newpage

\begin{references}
\bibitem[1]{serot}B. D. Serot and J. D. Walecka, Adv. Nuc. Phys. {\bf
16},

1 (1986).
\bibitem[2]{horowitz}C. J. Horowitz and B. D. Serot, Nucl. Phys. {\bf

A368}, 503 (1981).
\bibitem[3]{price1}C. E. Price and G. E. Walker, Phys. Rev. {\bf
C36}, 354

(1987).
\bibitem[4]{price2}R. J. Furnstahl, C. E. Price and G. E. Walker,
Phys.

Rev. {\bf C36}, 2590 (1987).
\bibitem[5]{odda}R. J. Furnstahl and C. E. Price, Phys. Rev. {\bf
C40},

1398 (1989).
\bibitem[6]{magnetic}J. A. McNeil, R. D. Amado, C. J. Horowitz, M.
Oka, J.

R. Shepard and D. A. Sparrow, Phys. Rev. {\bf C34}, 746 (1986).
\bibitem[7]{nearclosed}J. R. Shepard, E. Rost, C. -Y. Cheung, and J.
A.

McNeil, Phys. Rev. {\bf C37}, 1130 (1987).
\bibitem[8]{furnstahl}R. Furnstahl, Phys. Lett. {\bf 152B}, 313
(1985).
\bibitem[9]{shepard}J. R. Shepard, E. Rost, and J. A. McNeil, Phys.
Rev.

{\bf C40}, 321 (1989).
\bibitem[10]{mcneil}J. A. McNeil, R. J. Furnstahl, E. Rost, and J. R.

Shepard, Phys. Rev. {\bf C40}, 399 (1989).
\bibitem[11]{2loop}R. J. Furnstahl, R. J. Perry, B. D. Serot,  Phys.
Rev.

{\bf C40}, 321 (1989).
\bibitem[12]{ji}X. Ji, Phys. Lett. B208, 19 (1988).
\bibitem[13]{cohen}T. Cohen, in Workshop Procedings: From Fundamental

Fields to Nuclear Phenomena, p. 18, eds. J. A. McNeil and C. E.
Price,

World Scientific, Singapore (1991).
\bibitem[14]{prakash}M. Prakash, P. J. Ellis, and J. I. Kapusta,
Phys.

Rev. Lett. {\bf 45}, 2518 (1992).

\bibitem[15]{qcd} T. D. Cohen, Phys. Rev. Lett. {\bf 62}, 3027
(1989).
\bibitem[16]{chin}S. A. Chin, Ann. of Phys. (N.Y.){\bf 108}, 301
(1977).

\bibitem[17]{kurasawa}H. Kurasawa and T. Suzuki, Nuc. Phys. {\bf
A445}, 685 (1985).
\bibitem[18]{brockmann}R. Brockmann and J. Frank, Phys. Rev. Lett.
{\bf

68}, 1830 (1992).
\bibitem[19]{allendes}M. P. Allendes and B. D. Serot, Indiana
University

preprint, IU/NTC 92-04.


\end{references}
\end{document}